\documentclass[twocolumn,prb,aps,showpacs,preprintnumbers,amsmath,amssymb,superscriptaddress]{revtex4}
\usepackage{graphicx}
\usepackage{bm}
\begin{document}

\title{Charge and spin stripe in La$_{2-x}$Sr$_{x}$NiO$_{4}$ ($x=\frac{1}{3},\frac{1}{2}$)}
\author{Susumu Yamamoto}
\affiliation{Core Research for Evolutional Science and Technology, Japan Science and Technology Corporation (CREST-JST), Japan}
\author{Takeo Fujiwara}
\affiliation{Center for Research and Development of Higher Education, The University of Tokyo, Tokyo 113-8656, Japan}
\affiliation{Core Research for Evolutional Science and Technology, Japan Science and Technology Corporation (CREST-JST), Japan}
\author{Yasuhiro Hatsugai}
\affiliation{Institute of Physics, University of Tsukuba, 1-1-1 Tennodai, Tsukuba, Ibaraki 305-8571, Japan}

\begin{abstract}
Electronic structure of stripe ordered La$_{2-x}$Sr$_{x}$NiO$_{4}$ is
investigated.
The system with $x=\frac{1}{3}$ is insulator, in LSDA+U calculations,
and shows charge and spin stripe, consistent with the experimental results.
Highly correlated system of $x=\frac{1}{2}$ is studied
by using exact diagonalization of multi-orbital many body Hamiltonian
derived from LDA calculations and including on-site and inter-site Coulomb interactions.
The fluctuation of the residual spin on Ni$^{3+}$ (hole) site couples 
with the charge fluctuation
between Ni$^{3+}$ and Ni$^{2+}$ states 
and this correlation lowers the total energy.
The resultant ground state is insulator with charge and spin stripe 
of the energy gap 0.9~eV, consistent with observed one.
The on-site Coulomb interaction stabilizes integral valency of each Ni ion 
(Ni$^{3+}$ and Ni$^{2+}$), but does not induce the charge order. 
Two quantities, inter-site Coulomb interaction and anisotropy of hopping integrals, 
play an important role to form the charge and spin stripe order 
in a system of $x=\frac{1}{2}$.
\end{abstract}
\date{\today}
\maketitle

\newcommand{\STACK}[2]{\genfrac{}{}{0pt}{1}{#1}{#2}}

\section{Introduction}\label{sec_intro}

The stripe order of charge and spin has been found 
in several layered perovskites~\cite{Tranquada_1995,Tranquada_1994,Kimura_2001} 
and organic conductors.~\cite{Miyagawa_2000,Takano_2001}
Both of them are of pseudo two-dimensional (2D) electron system  
and have strong Coulomb interactions compared to hopping integrals. 
A single band system can be a model of organic conductors, 
while, in layered perovskites, a number of relevant orbitals depends on filling, 
crystal field, exchange splitting {\it etc}. 
The inter-site Coulomb interaction is essential to the
charge stripe in organic conductors and, then, extended Hubbard model is adopted
to explain the mechanism of stripe order there.~\cite{Seo_Hotta_Fukuyama_Review}
In layered perovskites, however, an origin of the stripe order is still controversial
particularly in the perovskites other than cuprate,
while the order in cuprate is attracting much attention in conjunction with
marked suppression of $T_{\rm c}$ with hole doping $\frac{1}{8}$ and rich physics.

Nickel compound La$_{2-x}$Sr$_{x}$NiO$_{4}$ (LSNO) is a typical system of
static stripe order of charge 
and spin.~\cite{Sachan_1995,Tranquada_1996,Yoshizawa_2000,Kajimoto_2003}
It is an insulator with the total spin $S=0$ in wide range of Sr doping $x$
($ 0 \leq x \leq 0.9$).~\cite{Cava_1991}
This stability of insulating phase is quite different from the variety of the phases 
in cuprate case; cuprate changes to metal, insulator and superconductor 
depending on the hole concentration. 
LSNO at $x=\frac{1}{3}$ shows the highest spin order temperature $T_{\rm SO} \sim 200$K,
because stripe order of charge and spin is commensurate with lattice periodicity.
The periods of charge and spin stripe are not generally commensurate 
to the lattice. 
Incommensurability $\epsilon$ increases with increasing $x$ and 
satisfies $\epsilon \gtrsim x$ in the region $\frac{1}{3}>x>0$, 
$\epsilon \lesssim x$ in the region $x > \frac{1}{3}$. 
Here, incommensurability $\epsilon$ is defined as displacement of peak positions 
of super structure from reciprocal lattice points.
Increase of $\epsilon$ saturates in the region of $\frac{1}{2}>x>\frac{1}{3}$ 
with the value $\epsilon\sim 0.44$.~\cite{Yoshizawa_2000}
In the region $x \gtrsim \frac{1}{2}$,
there exists commensurate charge ordered phase without magnetic order,
called a checkerboard type charge order,
between $T_{\rm SO}=80$K and $T_{\rm CO}=480$K.~\cite{Kajimoto_2003}
 Another experimental fact in  LSNO is the dependence of
the ratio between two lattice constants along with
$c$- and $a$-axis ($c/a$-ratio) on $x$.
The observed $c/a$-ratio has a maximum value at $x=\frac{1}{2}$, 
because holes are first doped into the $x^2-y^2$-orbital in a region $x<\frac{1}{2}$
and, then, additional holes are doped into the
$3z^2-1$-orbital when $x>\frac{1}{2}$.~\cite{Cava_1991}

We present two issues on LSNO in this paper by using both LSDA+U method 
and the exact diagonalization of many body Hamiltonian.
The first issue is that the inter-site Coulomb interaction is essential to
static charge order in doped layered perovskites.
Not only Hartree energy but also correlation energy 
due to the inter-site Coulomb interaction 
(beyond Hartree-Fock approximation) is important.
Because electron configurations fluctuate between hole (Ni$^{3+}$)
and non-hole (Ni$^{2+}$) states by hopping, 
the correlation energy is maximized at $x=\frac{1}{2}$.
In this situation, we need to diagonalize the many-body Hamiltonian
to know the true ground state at $x=\frac{1}{2}$.
The second issue is that the spin stripe order in LSNO 
is attributed to the structure of multi-orbitals.
A spin moment on a hole site is strongly correlated
with surrounding spin moments on non-hole sites 
in the multi-orbital system with fractional occupation,
which is a essential difference from the case of single-orbital systems.

The stripe order in layered perovskites is, in some cases, 
attributed to the Jahn-Teller (JT) distortion.~\cite{Zaanen_1989,Hotta_2004} 
In the present system, however, this is not the case, because the JT distortion is not
consistent with the local symmetry of the observed stripe order
of charge and spin in $x=\frac{1}{3}$ LSNO.
In the E$_{g}$ type JT distortion coupling with the ${\rm e}_g$ orbitals,
two oxygen atoms at opposite positions, centering Ni site, displace in opposite directions
with each other.
Consequently, the periodicity should be doubled along with -Ni-O-$\cdots$-O-Ni- line.
It contradicts to the observed tripled structure.
Distortion with ungerade mode can be consistent with the observed order
but increases the total energy.
The spin structure at $x=\frac{1}{2}$ is inconsistent with the JT distortion, too.

There is another candidate for the origin of stripe order in layered perovskites.
That is long-ranged Coulomb interaction.~\cite{Emery_1993}
Also it is well known that the inter-site Coulomb interaction 
stabilizes the charge order in quarterly filled (single-orbital) 
extended Hubbard model.~\cite{Ohta_1994,Seo_Hotta_Fukuyama_Review}
Thus, the inter-site Coulomb interaction can
stabilize the electronic structure of insulator with static stripe order.
In the electronic structure of LSNO, 
degeneracy of 3d ${\rm e}_g$-orbitals causes two important parameters:
splitting between $3z^2-1$-orbital and $x^2-y^2$-orbital 
and the hopping integrals.
It is reported that different values of these variables bring the system
different order.~\cite{Hotta_2004,Raczkowski_2006}
We should get reliable values of these parameters from the first principles
electronic structure calculations.

The present paper is organized as follows. 
The charge and spin stripe of LSNO ($x=\frac{1}{3}$)
is discussed in Sec.~\ref{sec_x_one_third}, based on  LSDA+U calculation. 
We show in Sec.~\ref{sec_x_half} the problem that charge ordered solution is not
found in LSDA+U calculations. 
Then we explain that the LSNO of $x=\frac{1}{2}$ 
is highly correlated system and the correlation energy in LSDA+U method is 
not enough to stabilize the charge ordered solution as a ground state.
In later Sections, exact diagonalization of many body Hamiltonian
is employed to investigate the electronic structure of LSNO ($x=\frac{1}{2}$).
The Hamiltonian derived from LDA calculation is explained 
in Sec.~\ref{sec_x_half_Hamiltonian}.
Section \ref{sec_x_half_charge_order} is devoted to  discussion of 
the charge stripe order and  charge correlation functions 
in the system of $x=\frac{1}{2}$ .
The excitation spectra and the energy gap of the system with $x=\frac{1}{2}$ is discussed
in Sec. \ref{sec_x_half_excitation}.
The spin stripe order of LSNO ($x=\frac{1}{2}$) is discussed 
in Sec.~\ref{sec_x_half_spin_order}. 
Finally Sec.~\ref{sec_conclusion} is the conclusion.

\section{Charge and spin stripe of LSNO ($x=\frac{1}{3}$) by using LSDA+U method}
\label{sec_x_one_third}

LSDA+U method~\cite{Anisimov_1991,Liechtenstein_1995}
in conjunction with the linear muffin-tin orbital method
with the atomic sphere approximation~\cite{Andersen_1975,Andersen_1984}
includes on-site Coulomb and exchange interaction $U,J$
with rotational invariant form.~\cite{Liechtenstein_1995,Oles_1983}
This on-site Coulomb term in LSDA+U Hamiltonian is called a ``Hubbard correction term''.
LSDA+U also includes the inter-site Coulomb interaction 
by means of Hartree energy.
Therefore, LSDA+U method can explain the physics of any charge ordered system,
if Hartree energy is enough to describe them.

For low doping systems ($x=0, \frac{1}{3}$) where the correlation energy 
induced by the inter-site Coulomb interaction is small,
the results of LSDA+U calculations can explain well experimental results.
Details of the calculations are as follows. 
The values of $U$ are chosen to be $7.5$~eV for Ni$^{2+}$ ions~\cite{Bocquet_1992a} 
and to be $7.0$~eV for Ni$^{3+}$ ions.~\cite{Imada_Fujimori_Tokura_Review}
The value of $J$ are chosen to be $0.88$~eV.
These values are consistent with those of constrained LDA.~\cite{Anisimov_1991}
In the present system,  NiO$_6$ octahedra are elongated~\cite{Cava_1991} 
with the direction of $c$-axis
and tilted.~\cite{Rodriguez-Carvajal_1991}
The elongation splits the two ${\rm e}_g$-orbitals by $\Delta=0.97$~eV 
in LDA calculation, 
where we denote the energy splitting between 
$3z^2-1$-orbital and $x^2-y^2$-orbital as $\Delta$.
On the other hand, the tilt of NiO$_6$ octahedra does not 
change the electronic structure in both LDA and LSDA+U calculations;
the localized magnetic moments on the Ni ion are 1.56~$\mu_B$ (with tilt) 
and 1.54~$\mu_B$ (without tilt), 
and the energy gap 3.66~eV (with tilt)  and 3.73~eV (without tilt).
Therefore, the tilt of NiO$_6$ octahedra is neglected hereafter.
We, then, fix the total volume of unit cell and 
set the $c/a$-ratio to be the observed value 3.26.
Crystallographic coordinates of atoms are fixed at those of $x=0$ and,  
with changing $x$, positions of all atoms are scaled. 
The magnetic unit cell at $x=0$
is a $\sqrt{2} \times \sqrt{2} \times 1$ supercell, and a $m
\sqrt{2} \times \sqrt{2} \times 1$ supercell at $x=\frac{1}{m}$.

\begin{figure}
    \begin{center}
    \includegraphics[width=8.5cm]{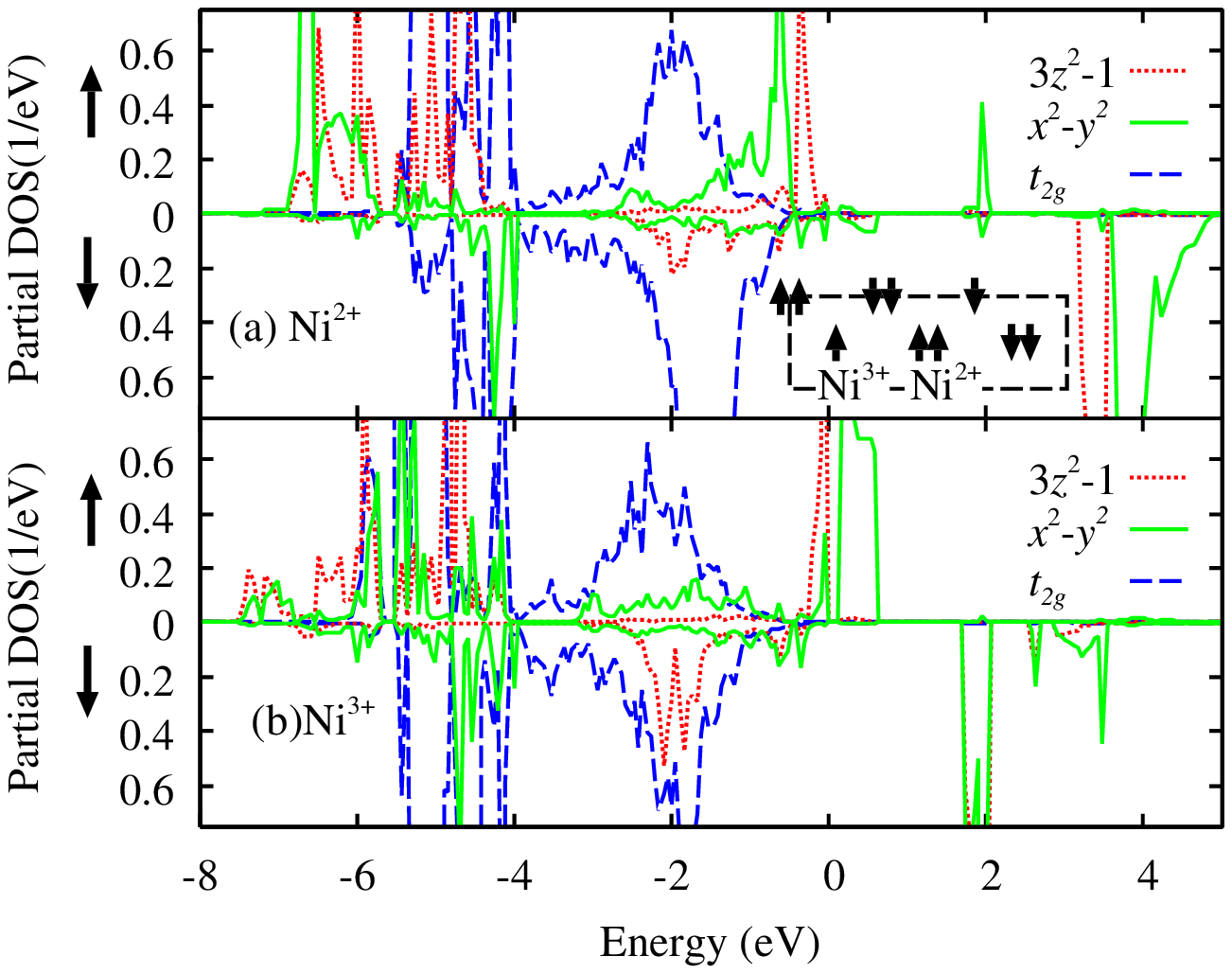}
    \caption{(color online) Projected density of states for each spin 
    (up and down indicated by arrows) of electron in
    La$_{\frac{5}{3}}$Sr$_{\frac{1}{3}}$NiO$_{4}$  by LSDA+U method for 
    (a) Ni$^{2+}$ and 
    (b) Ni$^{3+}$. Energy zeroth is fixed at the top of the occupied bands.
    Inset shows the location of
    each site. Local magnetic moment of each site is
    $-1.46$, $-0.94$, $1.51$, $-1.51$, $0.93$, $1.46\mu_B$, 
    respectively from the left to the right.}
    \label{fig_pdos_x33}
  \end{center}
\end{figure}

At $x=0$, calculated values of the band gap ($3.73$~eV) and 
local magnetic moment ($1.54~\mu_{\rm B}$) 
agree well with the  observed values of 4~eV~\cite{Ido_1991} 
and $1.68~\mu_{\rm B},$~\cite{Rodriguez-Carvajal_1991} respectively.
Figure~\ref{fig_pdos_x33} shows the projected density of states of LSNO 
at $x=\frac{1}{3}$.
The position of this $x^2-y^2$-orbital on the Ni$^{3+}$ site 
shifts to energy region just above the Fermi energy.
The calculated energy gap is $0.10$~eV.
The resultant spin structure is such that antiferromagnetic domains are separated by hole stripe,
and the magnetic moment localized on each site is 
$-1.46$, $-0.94$, $1.51$, $-1.51$, $0.93$, $1.46$~$\mu_B$, respectively, 
from the left to the right.
The center of the hole stripe is located on the Ni$^{3+}$ ions and 
no neighboring Ni$^{3+}$ ion exists.
The spins on two Ni$^{2+}$ ions sharing the same neighboring Ni$^{3+}$ ions 
are anti-parallel to each other (inset of Fig.~\ref{fig_pdos_x33}(a)).
Introduction of the multiple Slater determinant 
decreases the spin moments on Ni$^{3+}$ ions,
because the spin configuration with the opposite spin direction on Ni$^{3+}$ ions
gives the same energy with the present spin configuration.
This order is consistent with the experimentally observed stripe 
and should be assigned to the real ground state.
However, the calculated lowest energy state is of no charge order and 
the energy is lower by $0.5 \, {\rm eV}/{\rm cell}$ than that of the real ground state 
in Fig.~\ref{fig_pdos_x33}.
The spin ordered alignment of this calculated lowest energy state is 
different from that 
of the charge ordered state shown in the inset of Fig.\ref{fig_pdos_x33}(a). 
The local magnetic moment of the calculated lowest energy state is, on each Ni site 
in Fig.~\ref{fig_pdos_x33}(a),
$1.48$, $1.33$, $1.48$, $-1.48$, $-1.33$, $-1.48\mu_{\rm B}$, from the left to the light.
There are two problems responsible for the energy increase of the real ground state 
with the charge order.
One is the absence of correlation energy arising from the charge fluctuation 
between Ni$^{2+}$
and Ni$^{3+}$ configuration.
The other is the fact that LSDA+U method often underestimates the correlation energy 
of antiferromagnetic bonds.

\section{Electronic structure of LSNO ($x=\frac{1}{2},1$) by using LSDA+U method}
\label{sec_x_half}

\begin{figure}
    \begin{center}
    \includegraphics[width=8.5cm]{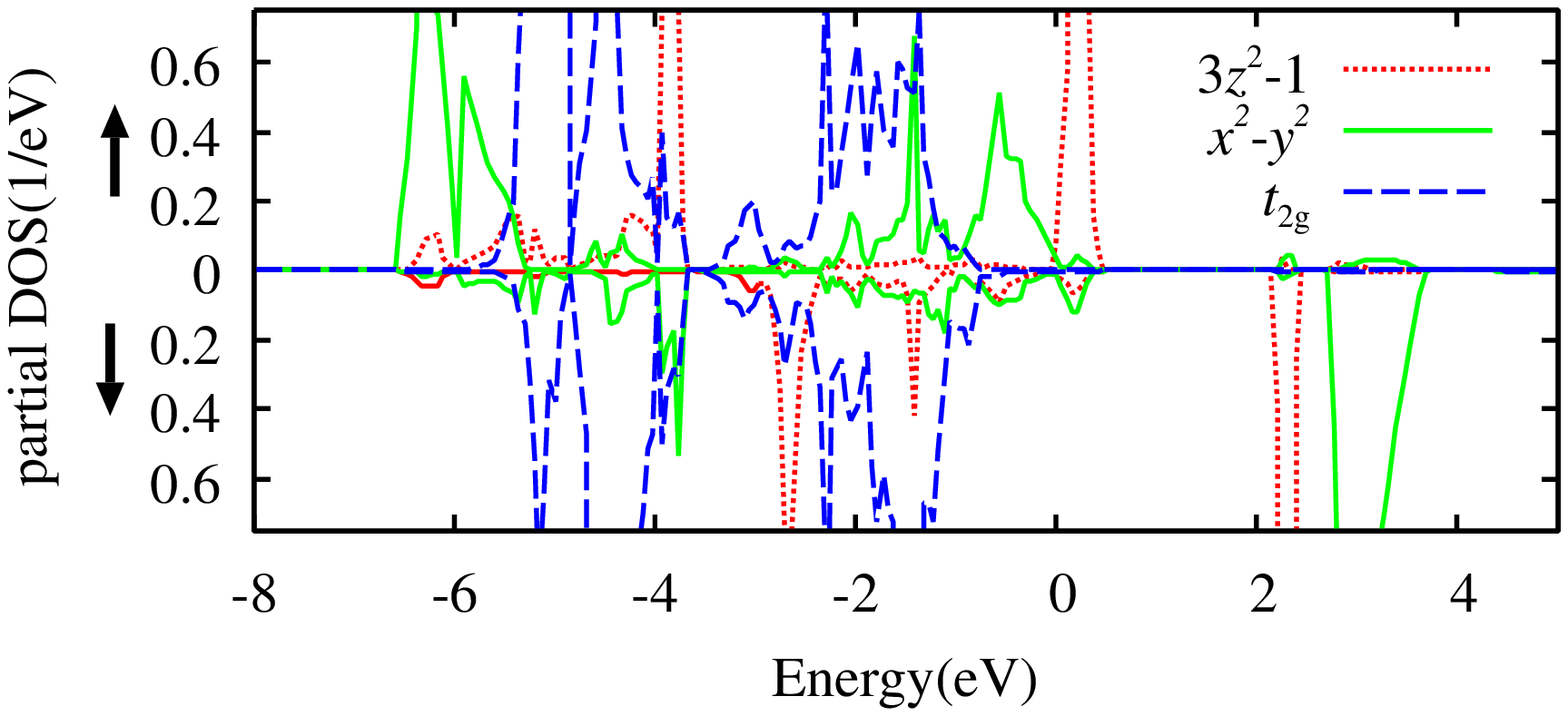}
    \caption{(color online) Projected density of states for each spin 
    (up and down indicated by arrows) of electron 
    in antiferromagnetic metallic LSNO($x=1$) by LSDA+U method.
    Energy zeroth is fixed at the Fermi energy.
	Resultant electronic structure is antiferromagnetic metal (see text).
	The hole states are 3$z^2-1$-orbital in contrast to $x=\frac{1}{3}$ case.
    Localized magnetic moment on each Ni site equals to $1.02~\mu_B$}
    \label{fig_pdos_x1}
  \end{center}
\end{figure}

LaSrNiO$_4$ ($x=1$) is observed to be paramagnetic metal, 
the occupation of the Ni ion distributes homogeneously and each ion is Ni$^{3+}$.
On the contrary, calculated ground state is antiferromagnetic metal and 
the projected density of states is shown in Fig.~\ref{fig_pdos_x1}. 
This contradiction between the observation and the calculation is 
due to unrealistic stabilization energy of
the magnetically ordered state in LSDA+U calculation of large on-site
Coulomb interaction $U,~J$, against the paramagnetic state. 
If we assume that the ground state is represented by multiple Slater determinants,
paramagnetic state would be represented as a linear combination of 
random spin configurations.
Then spin polarization within atomic sphere lowers the on-site Coulomb and 
exchange interaction energy and correlation energy stabilize the paramagnetic metal phase.
However, in the LSDA+U method, the ground state wavefunction is represented 
by a single Slater determinant.
Consequently, the paramagnetic state is only possible in the state 
where all ions have no spin polarization. 
This increases the total energy much, coupling with large $U,~J$.
A dominant component of the real ground state at $x=1$ should be
a linear combination of Slater determinants which have a single electron
on each site and spins are not ordered.

The LSDA+U calculation fails to present the paramagnetic metal phase as the 
ground state at $x=1$, 
but the correlation effects cause correct change of the splitting 
of two e$_g$ bands. 
Figure~\ref{fig_pdos_x1} shows that the $3z^2-1$ and $x^2-y^2$ bands locate 
above and below the Fermi energy, respectively.
Comparing  Figs.~\ref{fig_pdos_x33} and \ref{fig_pdos_x1},
one can see the change of the hole characters 
from the case of $x=\frac{1}{3}$ to that of $x=\frac{1}{2}$. 
In the $x=\frac{1}{3}$ system, the hole is doped in $x^2-y^2$-orbital,
while it is doped in $3z^2-1$-orbital in the $x=1$ system. 
This change of the hole character agrees with the experimental results 
described  in Sec.~\ref{sec_intro}. 
The origins of this change are the strong on-site Coulomb interaction $U$
between two orbitals and the difference of the dispersion of two orbitals/bands. 
In the LDA calculation of the system of $x=0$,
two ${\rm e}_g$ bands cross the Fermi energy.
The $3z^2-1$-band is narrow and lying 
on the energy region $-0.62$~eV$\sim$$0.69$~eV, 
and the $x^2-y^2$-band is wide and lying 
on the energy region $-0.86$~eV$\sim$$2.29$~eV, 
where the energy zeroth is fixed at the Fermi energy. 
When small amount of hole is introduced, 
the hole is doped into $x^2-y^2$-band, because $x^2-y^2$-band is averagely located
higher than $3z^2-1$-band. 
At the hole concentration $x=1$, orbital polarization is maximized if
only one band is occupied and the other is push up over the Fermi energy.
Because the lowest energy levels of ${\rm e}_g$-band 
are mainly consist of $x^2-y^2$-orbitals, 
the $x^2-y^2$ orbital is preferable to be occupied.
Thus, the hole character changes from $x^2-y^2$-orbital to $3z^2-1$-orbital at $x=1$. 
In LSDA+U calculation, however the spin and orbital polarization 
could not be completely maximized and the system becomes metal.
The potential correction in Table~\ref{tab_poco} shows this relation between 
the hole character and the strong on-site Coulomb interaction $U$.
Potential corrections are the derivatives of Hubbard correction term 
of LSDA+U Hamiltonian with respect to each element of local occupation matrix 
$\left<\hat{c}^{\sigma\dagger}_\alpha\hat{c}^\sigma_\beta\right>$ on each atom,
where $\alpha$ and $\beta$ denote atomic orbitals and $\sigma$ spin.
In the present system, potential corrections are diagonal matrices due to symmetry
and only diagonal elements are listed in Table~\ref{tab_poco}. 
In such case, each diagonal elements shows the energy level shift of 
respective orbital induced by on-site Coulomb interaction. 
Potential correction of $3z^2-1\uparrow$ ($x=\frac{1}{3}$)
nearly equals to that of $x^2-y^2\uparrow$ ($x=1$) and 
that of $x^2-y^2\uparrow$ ($x=\frac{1}{3}$) nearly equals to that of 
$3z^2-1\uparrow$ ($x=1$).
Therefore, the on-site Coulomb interaction applied to the
narrow $3z^2-1$- and wide $x^2-y^2$-band  induces
the difference of the hole character
between the low concentration ($x=\frac{1}{3}$) case
    and the high concentration ($x=1$) case.


\begin{figure}
	\begin{center}
    \includegraphics[width=8.5cm]{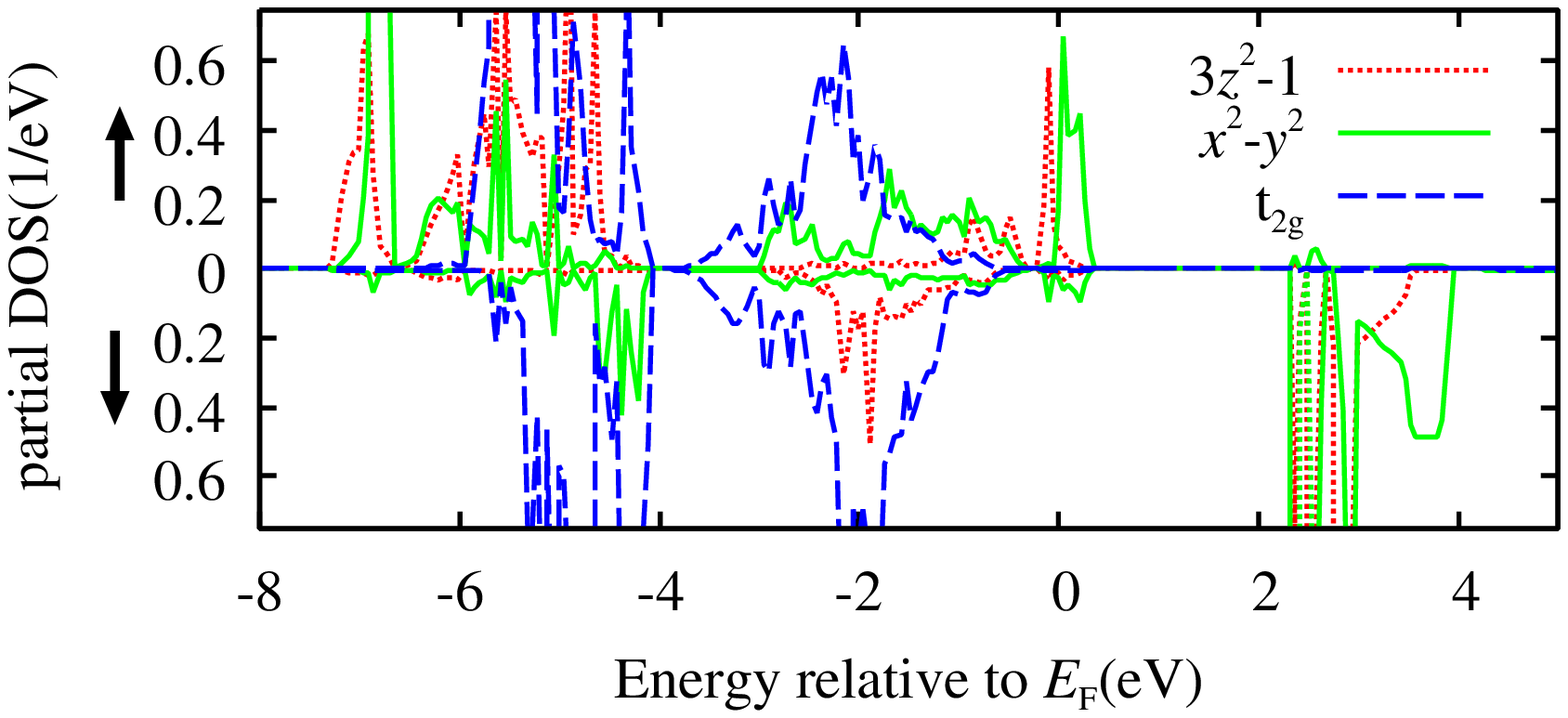}
	\caption{(color online) Projected density of states for each spin 
    (up and down indicated by arrows) of electron in
    antiferromagnetic metallic LSNO($x=\frac{1}{2}$) by LSDA+U method.
    Energy zeroth is fixed at the Fermi energy.
	Resultant electronic structure is antiferromagnetic metal (see text).
    Localized magnetic moment on each Ni site equals to $\pm 1.3$~$\mu_B$}
	\label{fig_pdos_x05}
	\end{center}
\end{figure}

In high-doped $x=\frac{1}{2}$ case,
the real system has a incommensurate stripe type charge order, as is described
in Sec. \ref{sec_intro}.
The ground state of LSDA+U calculation shows, however, no charge order.
Consequently, the electronic structure of the calculated system is metallic 
as shown in Fig.~\ref{fig_pdos_x05}.

The charge order of the system of $x=\frac{1}{2}$ 
would be expected to be a commensurate checker-board type order,
because of following three reasons.
The first is that the periodic boundary condition forces the charge order
to be commensurate to the lattice, 
and once we assume the incommensurability $\epsilon$ 
to be a rational value $\frac{1}{2}$,
the resultant charge order is checker-board type.
The second is that in the temperature range $T_{\rm SO} < T <T_{\rm CO}$,
the checker-board type charge order is observed in the real system with $x=\frac{1}{2}$.
The third reason is that the positive value of inter-site Coulomb interaction 
stabilizes the checker-board type charge order.
We must notice that the inter-site Coulomb interaction 
is included in LSDA+U Hamiltonian by means of Hartree energy.

We could not find a converged charge ordered solution, even as a metastable state,
by using LSDA+U calculation.
Localized magnetic moment of the four Ni ions in the unit cell equal to
$1.29, 1.32, -1.29, -1.32 \mu_B$, respectively.
Therefore, all the Ni ions are of the same type,
allowing the difference of local magnetic moment by $3$\%.
Figure \ref{fig_pdos_x05} shows the partial density of state of each orbitals of Ni ions.

The Hartree energy may be insufficient to stabilizes the charge order.
In the real system, the electron localized on the Ni$^{2+}$ site has extending tail on
Ni$^{3+}$ site and the center of localized electron 
can be fluctuating among Ni$^{2+}$ and surrounding Ni$^{3+}$ sites,
in order to lower the correlation energy. 
Because the ground state wavefunction is represented by a single Slater
determinant in LSDA+U calculation, 
the fluctuation of the center of localized electron 
is neglected,  then the correlation energy is underestimated.
Moreover, in the Hartree energy,
the contribution of the tail to the charge on Ni$^{3+}$ site and
the contribution of the head to the charge on Ni$^{2+}$ site are treated
as if they were separate electrons.
Such treatment gives an additional self-interaction and causes the increase of 
the estimated energy.
Because the number of nearest neighbor (n.n.) pairs of Ni$^{2+}$ and Ni$^{3+}$ ions 
are maximized in the checker-board type charge ordered system,
above two reasons become serious for the increase of estimated energy.

Now the prescription for resolving these problems are follows.
We need to preserve the antisymmetry for exchanging of any pair of single electron wavefunctions
in order to remove self-interaction and need to employ as many configurations of
Slater determinants as possible in order to correctly estimate the correlation energy.
Thus we employ the exact diagonalization of many body Hamiltonian in following Secs.

\begin{table}[h]
	\begin{tabular}{c|cccc}
		               &$3z^2-1\uparrow$	&$x^2-y^2\uparrow$	&$3z^2-1\downarrow$	&$x^2-y^2\downarrow$\\
 \hline\hline
 ${\rm Ni}^{2+}(x=\frac{1}{3})$& -2.07 & -2.06 & 3.80 &	3.29	\\
 ${\rm Ni}^{3+}(x=\frac{1}{3})$& -1.61 & -0.18 & 2.92 &	2.10	\\
 ${\rm Ni}^{3+}(x=1)$	       & -0.01 & -1.67 & 2.56 &	2.92
\end{tabular}
	\caption{Diagonal elements of potential correction (in unit of eV).
    Potential correction is a derivative of Hubbard correction term in LSDA+U
	Hamiltonian with respect to respective elements of occupation matrix.
	That gives orbital dependent potential corresponding to fluctuations
	of orbital occupation.
	}
	\label{tab_poco}
\end{table}

\section{Double orbital extended Hubbard Hamiltonian
for  LSNO ($x=\frac{1}{2}$) based on LDA calculation}
\label{sec_x_half_Hamiltonian}

To understand the ground state at $x=\frac{1}{2}$,
we adopt the extended Hubbard Hamiltonian $\hat{H}$ of ${\rm e}_g$ electrons
on two-dimensional (2D) simple square lattice 
derived from results of LSDA+U calculations, 
and diagonalize this many body Hamiltonian exactly
by using the Lanczos method with the inverse-iteration method; 
\onecolumngrid
\begin{equation}
\hat{H} =
 \sum_{\scriptstyle i,j, \alpha,\beta, \sigma}
  t_{i \alpha j \beta}   \hat{c}^{\dagger}_{i \alpha \sigma}
                         \hat{c}_{j \beta \sigma}
+ \sum_{i, \alpha, \sigma}
  \varepsilon_{i \alpha} \hat{c}^{\dagger}_{i \alpha \sigma}
                         \hat{c}_{i \alpha \sigma}
+ \frac{1}{2} \sum_{\STACK{i,\alpha,\beta,}{\gamma,\delta,\sigma,\sigma'}}
U_{\alpha \beta \gamma \delta}
\hat{c}^{\dagger}_{i \alpha \sigma}\hat{c}^{\dagger}_{i \beta \sigma'}
\hat{c}_{i \delta \sigma'}         \hat{c}_{i \gamma \sigma}
+ \frac{V}{2}
\sum_{\STACK{\left<i,j\right>,\alpha,}{\beta, \sigma, \sigma'}}
\hat{c}^{\dagger}_{i \alpha \sigma} \hat{c}_{i \alpha \sigma} \hat{c}^{\dagger}_{j \beta \sigma'} \hat{c}_{j \beta \sigma'} ,  \label{eq_reduced_hamiltonian}
\end{equation}
\twocolumngrid
\noindent
where the braces $\left<\cdots\right>$ denotes the summation over n.n. pairs 
and the symbol $t$ Slater-Koster type hopping parameters.
The on-site energy $\varepsilon_{i \alpha}$'s are 
determined by LDA calculation at $x=0$~:~\cite{LDA-result}
$t_{dd\sigma} = -0.543$~eV, $t_{dd\delta} = 0.058$~eV for the n.n. pairs
and 
$\frac{1}{4}t'_{dd\sigma} + \frac{3}{4} t'_{dd\delta} = -0.018$~eV,
$t'_{dd\pi} = -0.023$~eV for the second n.n. pairs and 
$\Delta=\varepsilon_{3z^2-1}-\varepsilon_{x^2-y^2}=0.97$~eV.
The energy zero-th is set at the midst of $\varepsilon_{3z^2-1}$ and 
$\varepsilon_{x^2-y^2}$.
The matrix elements of the intra-atomic Coulomb interactions 
$U_{\alpha\beta\gamma\delta}$ are represented as functions of
averaged Coulomb and exchange interaction $U, J$,
by using the same expressions as in LSDA+U method.~\cite{Liechtenstein_1995}
The values of $U, J$ are chosen to be $7.5$~eV, $0.88$~eV, respectively, 
on all sites.
The value of the inter-site Coulomb interaction $V$ is chosen as $0.5$~eV, 
except explicit indication of the value of $V$. 
Later we will explain the reason of this choice, $V=0.5$~eV, in detail.

\begin{figure}[h]
\begin{center}
    \includegraphics[width=7.5cm]{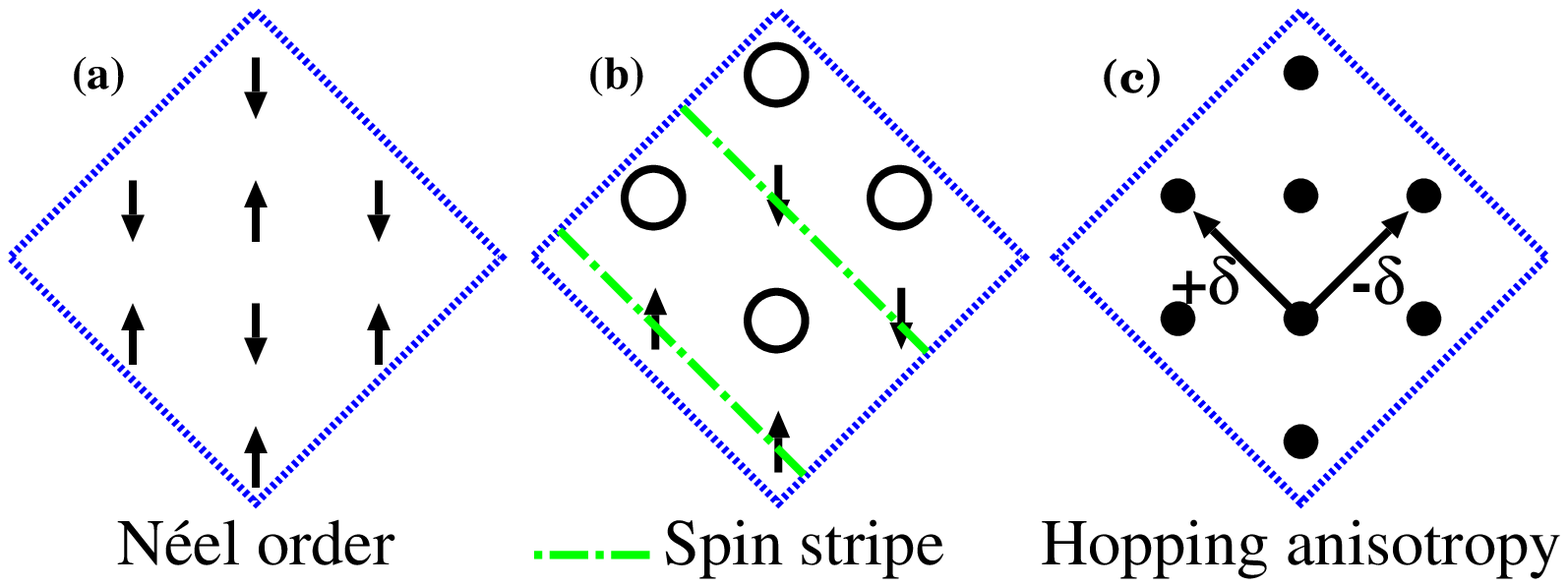}
    \caption{(color online) Two spin configuration
	at $x=\frac{1}{2}$ in a $ \sqrt{8} \times \sqrt{8} $ cell: 
       (a) N\'{e}el order and   
       (b) one of charge and spin stripe order among doubly degenerate ones. 
        Up and down arrows in (a) and (b) denote spins and circles in (b) denote holes.  
        In N\'{e}el order, spin
    correlation of four nearest neighbors has negative sign. 
        Those of two second nearest neighbors and a third nearest neighbor have
    positive sign. 
        In stripe order, two second nearest neighbors have opposite signs with each other.  
        (c) The hopping anisotropy of second nearest neighbors.
        }
\label{fig_spin_order}
\end{center}
\end{figure}

Because two diagonal directions are inequivalent in the real LSNO ($x=\frac{1}{2}$),
we introduce a parameter $\delta$ showing the anisotropy of the second n.n. 
(diagonal) hopping parameter $t'$.
We add to or subtract from $t'_{x^2-y^2,x^2-y^2}$ as $t'_{x^2-y^2,x^2-y^2}\pm \delta$,
depending on the direction, 
as depicted in Fig.~\ref{fig_spin_order}(c).
We investigate the $\delta$ dependence in the range of $ 0 \leq \delta \leq 0.02 $
so that the sign of the second n.n. hopping parameter does not change.
This anisotropy $\delta$ reserves translational symmetry.
As a result, finite value of $\delta$ does not induce charge order,
at least in the present range of $\delta$.
For example, the difference between the charge-charge correlation 
of the ground state at $V=0$~eV under $\delta=0.02$ and $\delta=0$ is at most $3$\%.
However, once charge order is induced by other quantities 
(in the present case, by inter-site Coulomb interaction $V$),
finite value of $\delta$ changes essentially the spin order of the system,
as is shown in Sec.~\ref{sec_x_half_spin_order}.
Finite value of $\delta$ neither changes essentially excitation spectra and energy gap.
The width of energy gap reduced by a few percent at $V=0.5$~eV.
Therefore, we fix the value of $\delta$ at $\delta=0$ 
in Secs.~\ref{sec_x_half_Hamiltonian}, 
\ref{sec_x_half_charge_order} and \ref{sec_x_half_excitation}.
We will change the value of $\delta$ in Sec.~\ref{sec_x_half_spin_order},
where the change of spin order induced by $\delta$ is discussed.

We diagonalize the many-body Hamiltonian of a system of 
twelve electrons on the planar $\sqrt{8} \times \sqrt{8}$ supercell, 
where each site has two ${\rm e}_g$-orbitals.
A periodic boundary condition is imposed, avoiding the bunching of electron at corners, 
and causes commensurate checker-board type order 
in the exact diagonalization result instead of incommensurate charge stripe
in real LSNO ($x=\frac{1}{2}$) as is mentioned in Sec.~\ref{sec_x_half}. 
The total ${\bf S}^2$ and $S_z$ of this system is invariant,
due to spherical symmetry of the spin-space.  
We then restrict the Hilbert space so as to total $S_z=0$,
which reduces the dimension of Hamiltonian to $(_{16}{\rm C}_{6})^2=64,128,064$.

\section{Charge order of LSNO ($x=\frac{1}{2}$)}
\label{sec_x_half_charge_order}

At first, we discuss the ground state of LSNO at $x=\frac{1}{2}$.
With increasing $V$, the ground state changes from 
singly degenerate state (State S) in the region $0{\rm eV}<V<0.41{\rm eV}$ 
to doubly degenerate (State D) $0.41{\rm eV}<V<1{\rm eV}$.
Smoothness of the connection among respective series of States S and D with respect to $V$
is discussed in Appendix~\ref{app_cont}.

\begin{figure}
    \begin{center}
    \caption{
	(color online) $V$ dependence of
	total energies and charge-charge correlations at $x=\frac{1}{2}$ and $\delta=0$,
	where $V$ is inter-site Coulomb	interaction.
	The state labeled ``S'' is singly degenerate while the other state
	labeled ``D'' is doubly degenerate. 
	Energy-zeroth is set at the ground state energy.
	State S is a ground state on the region $   0{\rm eV} < V < 0.41{\rm eV}$,
	State D is a ground state on the region $0.41{\rm eV} < V < 1{\rm eV}$.
	Suffix 1, 2, 3, 7 denotes following sites: origin, nearest neighbor, second nearest neighbor
	and third nearest neighbor, respectively.
    These site indexes are depicted in Fig.~\ref{fig_charge_order}(a).
    }
    \label{V_vs_corr}
  \end{center}
\end{figure}

\begin{figure}
	\begin{center}
    \includegraphics[width=8.0cm]{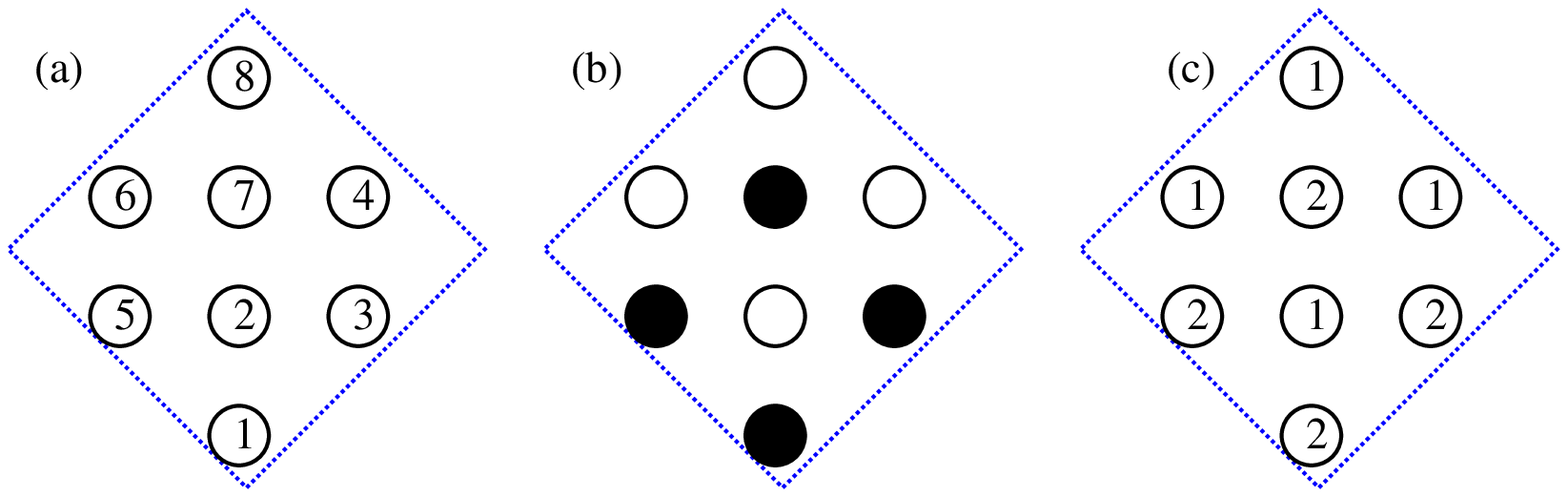}
	\caption{(color online) (a) Site index of the present system at $x=\frac{1}{2}$
	 in a $ \sqrt{8} \times \sqrt{8} $ cell.
	(b) A type of hole distribution of the Slater determinants with the complete checker-board type charge order
    (SDCCO in short), where closed and open circle denote non-hole and hole Ni site respectively.
	Another type of distribution is obtained by exchanging the hole site and non-hole site.
	(c) The occupation distribution of SDCCO in (b).
	A number in the open circle shows the occupation number of respective site.
	}
	\label{fig_charge_order}
	\end{center}
\end{figure}

Figure~\ref{V_vs_corr} shows the $V$ dependence of the total energies and 
the correlation functions of charge fluctuation 
$\left<\delta \hat{n}_i\delta \hat{n}_j\right>$, 
where $\delta \hat{n}_i=\hat{n}_i-\left<\hat{n}_i\right>$ and 
$\delta$ is fixed at $0$. 
At $V=0.5~eV$, charge correlations $\left<\delta \hat{n}_i\delta \hat{n}_j\right>$ 
of State D equal to $-0.135$, $0.093$, $0.096$ for n.n. ($(i,j)=(1,2)$),
second n.n. ($(i,j)=(1,3)$) and third n.n. ($(i,j)=(1,7)$) pairs respectively,
and this corresponds to checkerboard type charge order shown 
in Fig.~\ref{fig_charge_order}(b).
The checker-board type charge order in State D exists even at $V=0$eV, 
though the amplitude is very small.
The charge correlations for n.n., second n.n. and third n.n. pairs at $V=0$eV
equal to $-0.074, 0.007, 0.019$, respectively.
These values satisfy the checker-board type charge order:
The positive correlation between second n.n pair nearly equals to 
that between third n.n pair, 
and the correlation between the first n.n. pair is negative.
The charge correlations in State S are homogeneous 
in comparison with those in State D. 
The charge correlation in State S at $V=0$~eV of first, second and third n.n. pairs equal
to $-0.046,-0.013,-0.056$, respectively. 
Therefore, State S at $V=0$~eV does not have any charge order. 
The absence of charge order in State S in $0~{\rm eV}<V<0.4~{\rm eV}$
is attributed to a slow increase of charge correlations 
with increasing $V$ in this region of $V$.

Now we discuss three features in Fig.~\ref{V_vs_corr},
two are about correlation functions of State S and 
one about the energy difference between States S and D;
(i) Rapid increase of the charge correlation of State S 
in the region $0.4~{\rm eV}<V<0.6~{\rm eV}$, 
(ii) merging charge correlation of State S into that of State D, 
in the region $V>0.6~{\rm eV}$ and 
(iii) the constant energy difference between States S and D 
in the region $V>0.6~{\rm eV}$. 
All these features are related to respective wavefunctions of States S and D.

Let us define the ideal complete checker-board type charge order whose Slater determinant
wavefunction (SDCCO) should satisfy the following two conditions, (a) and (b),  
then we specify the characteristics of real wavefunctions of States S and D by using the
overlap with SDCCO's:
(a) every site has occupation one or two, and 
(b) all neighboring sites of a singly occupied site 
are doubly occupied sites and vice versa.
A type of their distribution of occupation is shown in Fig.~\ref{fig_charge_order}~(c). 
Assuming both transfer integral and on-site exchange parameter $J$ equal to zero, 
SDCCO's would be the ground states, where all the spin configurations are degenerate,
because the numbers of neighboring hole (Ni$^{3+}$) and non-hole (Ni$^{2+}$) pairs are 
maximized.
In an actual calculation, transfer integrals are finite and the SDCCO's 
are not eigenstate. 
The value of $V$ determines how much SDCCO's are hybridized into respective eigenstates.
The more $V$ increase, the more hybridization of SDCCO's are preferable energetically.

\begin{figure}
    \begin{center}
    \caption{
	(color online) $V$ dependence of
	the weight of Slater determinants with complete checker-board type charge order
    (SDCCO in short), where $V$ is inter-site Coulomb interaction.
	The state labeled ``S'' and ``D'' are the same states as in Fig.~\ref{V_vs_corr}.
	$\delta=0$.
    }
	\label{fig_weight_SDCCO}	
  \end{center}
\end{figure}

In the region of $0~{\rm eV}<V<0.4~{\rm eV}$, the ground state is homogeneously extending 
State S since the kinetic energy is a source of gain of the total energy. 
On the other hand, in the region of $0.4~{\rm eV}<V$, the ground state changes to  
the charge ordered State D since the source of energy gain is the correlation 
energy due to the inter-site Coulomb interaction $V$.
The dependence of the total weight of SDCCO's is shown in Fig.~\ref{fig_weight_SDCCO}.
Comparing Figs.~\ref{V_vs_corr} and \ref{fig_weight_SDCCO}, 
the above features (i), (ii) and (iii) appear as follows. 
Since the wavefunction of the ground state State S is homogeneous, 
the overlap with SDCCO is small in the region of $0~{\rm eV}<V<0.4~{\rm eV}$.
We discuss more detail about the weight of SDCCO's of State S at $V=0$~eV
in Appendix \ref{app_W_SDCCO}.
Then, in the range $0.4~{\rm eV}<V<0.6~{\rm eV}$, 
the weight of SDCCO's rapidly increases.
Because both States S and D are eigenstates of $\hat{H}$ and orthogonal with each other,
the coefficients of SDCCO's in State S are different from those in State D.
In the range $0.4~{\rm eV}<V<1~{\rm eV}$,  States S and D have 
the same charge order but the spin order is quite different, 
which is discussed in Sec.~\ref{sec_x_half_spin_order}.
Then, the difference of the total energies between States S and D 
is attributed to the difference of spin configurations.
As is discussed in Sec.~\ref{sec_intro},  
$T_{\rm SO}<T_{\rm CO}$ in the real system with $x=\frac{1}{2}$ 
and, therefore, the energy scale of spin order is smaller 
than that of charge order. 
In fact, we can see a large energy difference between States S and D in the 
region of $0~{\rm eV}<V<0.4~{\rm eV}$, since the charge order 
different between States S and D here. 
Then , the characteristics of the charge order are the same 
in States S and D in the range $0.6{\rm eV}<V<1{\rm eV}$ (Feature (ii)), 
as seen in the charge correlation functions and the same weights of SDCCO's 
in States S and D, 
and the constant energy difference between States S and D (Feature (iii)). 
The observed high temperature state ($T>T_{\rm SO}$) in LSNO of $x=\frac{1}{2}$ 
might be a mixture of charge ordered eigenstates
with significant weight of SDCCO's, 
including the doubly degenerate States D and the singly degenerate State S.

As mentioned in Sec.~\ref{sec_intro},
the real ground state of LSNO ($x=\frac{1}{2}$) shows charge stripe order,
incommensurate to the lattice.
And as mentioned in Sec.~\ref{sec_x_half},
the periodic boundary condition forces the charge order to
be commensurate charge order of checker-board type.
Therefore, the ground state of the Hamiltonian should have
checker-board type charge order,
and we choose $V=0.5$~eV from now on.
The value $V=0.5$~eV is not very unrealistic because
a recent report shows $V=0.34$~eV in GW approximation of LaMnO$_3$.~\cite{Nohara_2006}

\section{Excitation spectra and energy gap of LSNO ($x=\frac{1}{2}$)}
\label{sec_x_half_excitation}

We show calculated single-particle spectral functions 
of the system with $x=\frac{1}{2}$ in Fig.~\ref{fig_spectral_function}, 
in order to see whether the spectra have energy gap and the system is insulator 
as the observed case.
Inter-site Coulomb interaction is fixed at $V=0.5$~eV as discussed 
in Sec.~\ref{sec_x_half_charge_order}. 
Three ${\bm k}$-points are chosen; 
${\bm k}=(0,0)\frac{1}{a}, 
(\frac{\pi}{4},\frac{\pi}{4})\frac{1}{a}$ and 
$(\frac{\pi}{2},0)\frac{1}{a}$,
where $a$ is the nearest Ni-Ni distance, 
the positions of Ni ions are $(n,m)a$, 
and the two translation vectors are $(-2,2)a$ and $(2,2)a$. 
Each ${\bm k}$-point corresponds to periodic or antiperiodic 
boundary condition along respective translation vector. 
In order to obtain continuous spectra, we introduce  
an imaginary energy $\eta$ of $0.01$~eV 
and the value of $\eta$ determines the resolution of the spectra. 
The value of $\eta$ must satisfies $\eta \gtrsim \frac{\mbox{total width of spectra}}{m}$, 
where $m$ denotes the $\mbox{the highest power of}\ \hat{H}$, 
because $m$ is the upper limit of the number of $\delta$-functions in the spectra. 
In the Lanczos method, the truncation error easily breaks down the orthogonality 
of Lanczos vectors. 
Because calculating spectral function may need the large $m$, 
we employ the shifted-COCG method,~\cite{Takayama_2006}
which is a variant of the conjugate gradient (CG) method. 
It must be noticed  that the shifted-COCG method is numerically stable.
Because shifted-COCG method requires an reference energy in the energy region
of a peak of spectral weight, we need to know rough profile of spectral weight. 
Therefore, the shifted-COCG method should be used 
only when we need very fine profile of spectral function. 
More details will be explained elsewhere.

In the present case, the ``total width of spectra'' is roughly estimated as 
$E({\rm Ni^{2+}})-E({\rm Ni^{3+}}) \sim U=7.5$eV for ionization levels,
and affinity levels, separately. 
Then we choose $m=800$.
We calculate the spectral function of $m=160$ by Lanczos method, and then,
calculate that of $m=800$ by the shifted-COCG method.
Actually, the width of spectra is  wider than $U$ due to the mixing of
higher/lower occupation configurations. 
The spiky structure of spectra is an artifact due to the choice of smaller $\eta =0.01$~eV.
Our choice of the value of $\eta$ is intending to show the gap structure between
ionization levels and affinity levels, at every ${\bm k}$-point. 
We fix the value of hopping anisotropy $\delta=0$. 
The hopping anisotropy $\delta$
does not change the shape of the spectral function in the range of $0 \le \delta<0.02$. 
The width of the gap at ${\bm k}=(0,0)\frac{1}{a}$ in Fig.~\ref{fig_spectral_function} 
is smallest among three ${\bm k}$-points, 
where the highest ionization level (HIL) is located at $9.4$~eV 
and the lowest   affinity  level (LAL) $10.3$~eV.
Therefore, we conclude that this system is insulator with an energy gap of $0.9$~eV.

\begin{figure}
\begin{center}
\caption{
(color online) Single-particle spectral function of 8 sites,
$N=12$ ($x=\frac{1}{2}$) system with $\delta=0$ and $V=0.5$~eV. 
Three panels are corresponding to respective ${\bm k}$-points,
$(\frac{\pi}{2},0)\frac{1}{a}$, $(\frac{\pi}{4},\frac{\pi}{4})\frac{1}{a}$, $(0,0)$,
from the top to the bottom.  
Each value of ${\bm k}$ corresponds to different boundary condition. 
Two arrows near $10$~eV shows the position of
    the highest ionization level  $9.4$eV 
and the lowest   affinity  level $10.3$eV in case of ${\bm k}=(0,0)\frac{1}{a}$.
Spectral functions are approximated by the polynomial of 800 degree in $\hat{H}$,
where $\hat{H}$ denotes the Hamiltonian. 
A small imaginary part ($0.01$~eV) is added to the energy $\omega$ for 
smearing $\delta$-function peaks.
}
\label{fig_spectral_function}
\end{center}
\end{figure}

It is important to know the symmetry of the single electron wavefunctions 
of HIL and LAL,
in order to show that the ``gap'' is not an artifact of discretized ${\bm k}$-point.
The symmetry of respective single electron wavefunction can be labeled 
by crystal momentum ${\bm k}$. 
If the ``gap'' is an artifact and the ground state is metal, 
HIL and LAL are labeled by different ${\bf k}$. 
Actually many levels overlap and form a single peak-like structure 
because of complicated interactions,
especially in the energy range near LAL.
Even the definition of the ``single electron wavefunction'' are not clear in
such a circumstance. 
Therefore, we define a ``single electron wavefunction'' as follows, here.
First we choose the bottom ($E_{\rm b}$) and top ($E_{\rm t}$) of a single
peak-like structure. 
Next we integrate the Green function of matrix form
in the energy range $E_{\rm b}<E<E_{\rm t}$:
\begin{equation}
	-\frac{1}{\pi} \int_{E_b}^{E_t} {\rm Im} G_{i\alpha j\beta}^R(\omega) {\rm d}\omega.
\label{partial_G}
\end{equation}
and then we diagonalize it. 
This matrix corresponds to energetically partitioned 
$\left< \hat{c}^\dagger_{i\alpha} \hat{c}_{j\beta} \right>$ ~ $(E_b<E_t<E_F)$ or
$\left< \hat{c}_{i\alpha} \hat{c}^\dagger_{j\beta} \right>$ ~ $(E_F<E_b<E_t)$.
Therefore, the eigenvector of above matrix corresponds to
the single electron wavefunction related to the peak-like structure 
and 
the eigenvalue corresponds to the occupation number of the single
electron wavefunction. 
We summarize in Table~\ref{tab_occ} the $E_{b}, E_{t}$, eigenvalues and 
the $\vec{k}$ representing the symmetry of eigenvectors for the peak-like structure near 
HIL and LAL.
At ${\bm k}=(0,0)\frac{1}{a}$,  HIL and LAL single electron wavefunctions share
the same symmetry. This can not occur in the metallic system.
Therefore, we can conclude that the split between HIL and LAL is not 
an artifact and the system is an insulator.

\begin{table}
\begin{tabular}{c|ccccc}
	& $E_{\rm b}$(eV) & $E_{\rm t}$(eV) &
 \begin{tabular}{c} Peak \\ area \end{tabular}&
{\renewcommand\arraystretch{0}
 \begin{tabular}{l} The largest \\ eigenvalue and \\ its degeneracy  \end{tabular}
}\vspace{1mm}&
{\renewcommand\arraystretch{0}
 \begin{tabular}{c} Symmetry of \\ eigenvectors \\ $\vec{k}$($\frac{1}{a}$) \end{tabular} 
}\vspace{1mm}\\
			\hline \hline
	& \multicolumn{3}{c}{at $\vec{k}=(0,0)\frac{1}{a}$ }\\
	HIL &  9.15 & 9.55 & 0.37 & 0.09$\times$4 & $(\frac{\pi}{2},\frac{\pi}{2})$ \\
	LAL & 10.05 &11.15 & 2.87 & 0.41$\times$4 & $(\frac{\pi}{2},\frac{\pi}{2})$ \\
	& \multicolumn{3}{c}{at $\vec{k}=(\frac{\pi}{4},\frac{\pi}{4})\frac{1}{a}$} \\
	HIL &  8.18 & 8.98 & 1.72 & 0.63$\times$2 & $(\frac{\pi}{4},\frac{\pi}{4})$ \\
	LAL & 10.02 &11.02 & 2.99 & 0.44$\times$4 & $(\frac{-\pi}{4},\frac{3\pi}{4})$ \\
	& \multicolumn{3}{c}{at $\vec{k}=(\frac{\pi}{2},0)\frac{1}{a}$} \\
	HIL &  8.18 & 9.18 & 1.88 & 0.38$\times$4& $(\frac{\pi}{2},0)$ \\
	LAL & 10.12 &11.12 & 2.90 & 0.42$\times$4& $(\frac{\pi}{2},\pi)$
	\end{tabular}
	\caption{
	Bottom and top of the energy of peak-like structure, area of the structure,
    the largest eigenvalue (and its degeneracy after ``$\times$'' symbol) related to 
    ``single electron wavefunction'' (see text) and 
    the representative $\vec{k}$ which shows the symmetry of the single electron wavefunction.
	In the right column, single $\vec{k}$ is written and the other equivalent ($\vec{k}$) are omitted.
    HIL and LAL stands for highest ionization level (HIL) and lowest affinity level (LAL) respectively.
    Each ${\bm k}$-points corresponding to the boundary conditions are written in the leading lines. $\delta=0$.
	}
	\label{tab_occ}
\end{table}

\begin{figure}
\begin{center}
\caption{
(color online)
Spectral functions of 8 sites,
$N=12$ ($x=\frac{1}{2}$) system are drawn at $\delta=0$.
(a)Each panel corresponds to $V=0.5$, $0.2$ and $0$~eV from the top to the bottom.
In order to show the overlap between 
the ionization levels and the affinity levels at $V=0$~eV,
they are drawn separately in the upper and the lower halves of respective panels.
Spectral functions are approximated by the polynomial of $\hat{H}^{160}$,
where $\hat{H}$ denotes the Hamiltonian. 
Imaginary energy $0.01$~eV is added to the energy $\omega$
for smearing $\delta$-function peaks.
Thin vertical line shows the highest ionization levels $9.4$~eV, $7.9$~eV and $6.9$~eV
(at $V=0.5$, $0.2$ and $0$~eV). 
(b)
Each orbital component ($3z^2-1$- or $x^2-y^2$- orbital) of the spectral function
is drawn separately, near the highest ionization level.
Thin vertical line shows the highest ionization levels, same as Fig.(a).
Both the highest ionization level and the lowest affinity level consist
of the $x^2-y^2$ orbital and the symmetry is labeled by
$\vec{k}=(\frac{\pi}{2},\frac{\pi}{2})\frac{1}{a}$. See also Table \ref{tab_occ}.
}
\label{fig_spectral_function_vs_V}
\end{center}
\end{figure}

Now we show that the inter-site Coulomb interaction $V$ induces the energy gap 
between occupied- and unoccupied- states, not on-site Coulomb interaction $U$.
Figure~\ref{fig_spectral_function_vs_V}~(a),~(b) shows the ``spectral function'' 
with respect to State D with $V=0.5$, $0.2$ and $0$~eV. 
Note that, since State D is not the ground state at $V=0.2$~eV and $V=0$~eV, 
they (for $V=0.2$~eV and $V=0.5$~eV cases) are not satisfying 
the definition of spectral function.
These spectral functions are approximated by rational function of the degree of 
$\hat{H}^{160}$ by using Lanczos method.
The top panel of Fig.~\ref{fig_spectral_function_vs_V}~(a) 
and the bottom panel of Fig.~\ref{fig_spectral_function} 
are the same, except that the latter uses higher degree of $\hat{H}^{800}$ 
and the shifted-COCG.
They are sharing characteristic peaks of HIL and LAL.
Therefore, 160 degree is enough to discuss the characteristics of spectra. 
At $V=0.5$~eV (top panel), HIL and LAL are $(x^2-y^2)$- and $(3z^2-1)$-orbitals, 
respectively, and this agrees well with experimental observation 
and the spectrum in Fig.~\ref{fig_pdos_x05}.
At $V=0.2$~eV (middle panel), HIL lowers to $7.9$~eV and the energy gap 
becomes smaller than that at $V=0.5$~eV. 
At $V=0$~eV (bottom panel), HIL lowers to $6.9$~eV and 
two excitation peaks overlap with each other in  HIL and LAL, 
{\it i.e} the energy gap vanishes.
Thus we can conclude that
the system with $x=\frac{1}{2}$ becomes insulator due to $V$, not $U$, 
and the inter-site Coulomb interaction $V$ causes 
the energy stabilization and the opening the gap of the ground state State D. 
The gap width in the bulk limit is discussed in Appendix \ref{app_gap}.

\section{Spin order of LSNO ($x=\frac{1}{2}$)}
\label{sec_x_half_spin_order}

\begin{figure}
 \begin{center}
 \includegraphics[angle=-90,width=8cm]{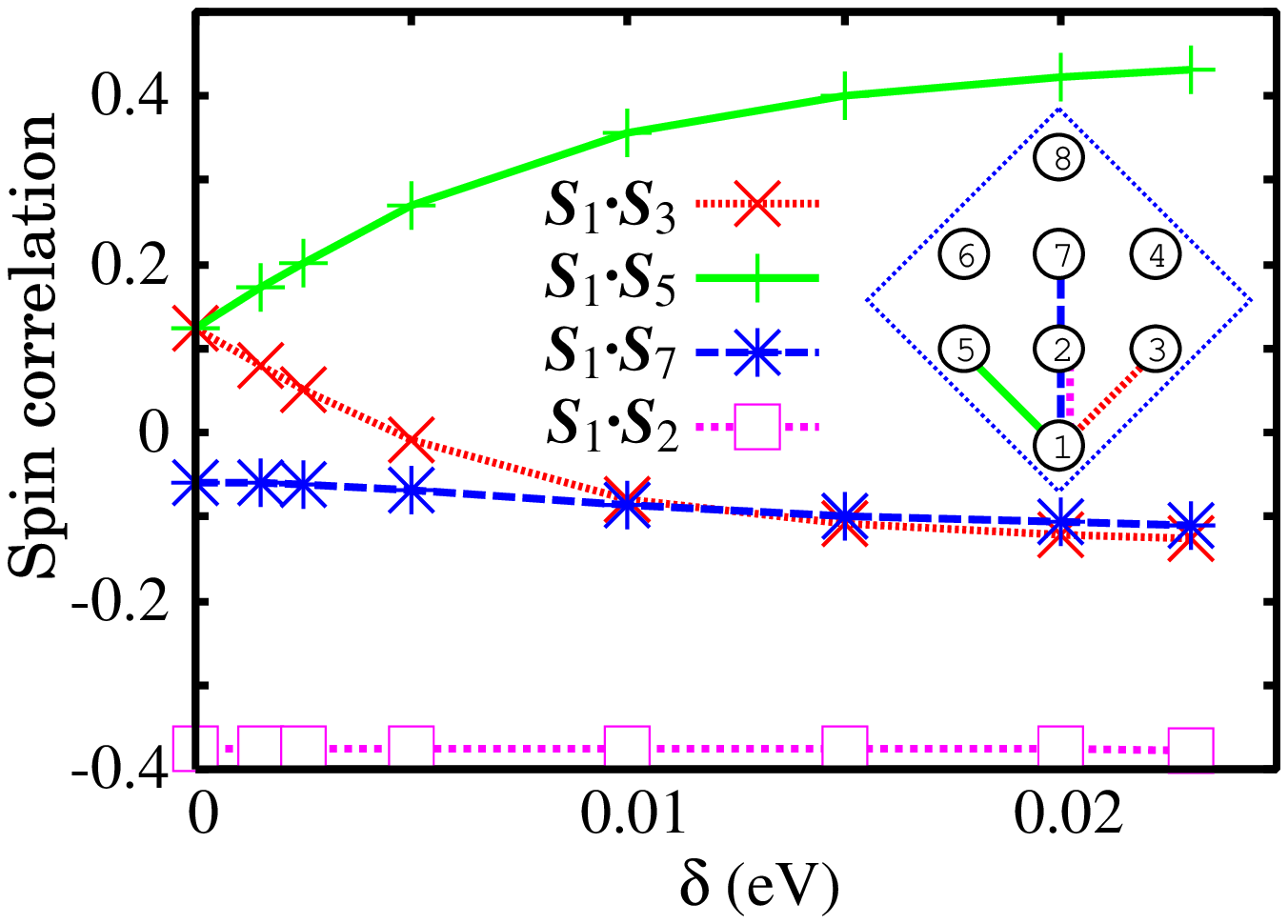}
 \caption{(color online) Spin correlation function $\left<{{\bf S}_i}\cdot{{\bf S}_j}\right>$  
 with the anisotropy parameter $\delta$ of
 the second nearest hopping integrals.
 The value of $V$ equals to $0.5$~eV.
 Inset shows site location.}
 \label{distortion-correlation_and_energy}
 \end{center}
\end{figure}

We discuss the spin stripe order induced by the anisotropy of hopping integral between the n.n. sites.
Introduction of the anisotropy does not contradict to the local symmetry of
observed charge and spin stripe. 
This anisotropy with $\delta$ may couple with orthorhombic lattice distortion
of ${\rm T}_{2g}$ symmetry.
Coupling to the distortion reminds us of JT mechanism.
However, the T$_{2g}$ distortion is not a JT mode for the ${\rm e}_g$ electron,
and therefore, this mechanism is not of JT mechanism.

The spin correlation function $\left<{{\bf S}_i}\cdot{{\bf S}_j}\right>$ at $\delta=0$
equals to $-0.37$, $0.12$, $-0.07$ respectively when $V=0.5$~eV 
and shows a lack of long range order.
Equivalence of two diagonal directions is an origin of the two-fold degeneracy,
as is shown in the result of ${\bm k}=(\frac{\pi}{4},\frac{\pi}{4})\frac{1}{a}$ 
in Sec.~\ref{sec_x_half_charge_order}.

Because large amount of holes are doped in $x^2-y^2$-orbital,
the electron in $x^2-y^2$-orbital on Ni$^{2+}$ can have a long tail
extending to Ni$^{3+}$ site.
The hopping part of the total energy per electron is $-0.30$~eV and
its absolute value is as twice large 
as that in the parent material ($x=0$, $N=16$), $-0.15$~eV.
Following the localized electron picture,
increase of hopping energy indicates larger hybridization of $x^2-y^2$-orbital
between Ni$^{2+}$ and Ni$^{3+}$.
Then, spin on Ni$^{2+}$ site has strong correlation with that on Ni$^{3+}$ site.
Thus the spin correlation in $x=\frac{1}{2}$ system is more complicated than
that in $x=0$ system where the localized spin picture is applicable.

The present parameter set in $\hat{H}$ is another origin of the complexity 
of spin structure. 
If $J \gg \Delta$, the majority spin is preferable in order to lower the exchange energy
when the electron visited the Ni$^{3+}$ site.
In contrast, if $\Delta \gg J$, the minority spin is preferable 
in order to lower on-site energy. 
The present values of $J$ and $\Delta$ satisfy neither of two conditions.
In addition to that, transfer integral and the inter-site Coulomb interaction $V$ 
has the same energy scale. 
This causes the competition between the extended state and the 
localized single electron wavefunction. 
All these competitions may cause the instability against the perturbation.

In the case of $\delta\neq 0$, the ground state energy decreases 
with increasing $\delta$ and can be expressed as ($36.75-104.3\delta^2$)~eV. 
The spin correlation function is depicted in Fig.~\ref{distortion-correlation_and_energy}. 
The value of the correlation function between nearest neighbors,
$\left<{{\bf S}_1}\cdot{{\bf S}_2}\right>=-0.37 \sim -0.38$, does not change.
The local spin moment is $\left<{{\bf S}_i}^2\right>=1.31$. 
If two electrons localized on one site in a triplet state,
four electrons are localized on the four n.n. sites,
and the total spin $S$ of this totally six electron system equals to 1,
then $\left<{{\bf S}_1}\cdot{{\bf S}_2}\right>=-0.375$,  which is almost the same as 
the calculated value of $\left<{{\bf S}_1}\cdot{{\bf S}_2}\right>$.
In the range of $\delta \gtrsim 0.01$~eV,
correlation functions of the two types of second neighbor pairs 
have opposite signs and the structure of the calculated spin stripe order
is consistent with the observed one.
This critical value $\delta=0.01$~eV is less than 2\% of
the absolute value of the n.n. hopping $t_{dd\sigma}$, and thus
small anisotropy changes the spin structure drastically.

The spin order is quite different between systems of the single-orbital 
and of the multi-orbitals.
The former on 2D square lattice is the N\'{e}el order on $\sqrt{2}\times\sqrt{2}$ cell,~\cite{Ohta_1994} 
and the latter is the charge and spin stripe.
Once we assume only $x^2-y^2$-orbital at each site and put 8 electrons on 16 sites, 
we get the value of $\left<{{\bf S}_1}\cdot{{\bf S}_2}\right>$ equal to $-0.03$
which is very small compared to the value $-0.38$ in LSNO.\cite{co_variance}
This reduction of $\left<{{\bf S}_1}\cdot{{\bf S}_2}\right>$ 
in the single-orbital system is attributed to the absence of localized spin 
on a hole site and consistent with the N\'{e}el order mentioned above.
However, a spin remains on a hole site in LSNO ($x=\frac{1}{2}$).
Since two neighboring sites of a hole site have spins of opposite direction
with each other, the spin state of a hole site is a linear combination of up- and down-spin
due to symmetry.
Because no term in the Hamiltonian flips a spin on a hole site, 
two antiparallel-spin electrons on two hole sites must exchange
their positions through hopping process, in order to create such a linear combination.
This hopping process lowers correlation energy and the ground state must be represented by the
multiple Slater determinant.
In all hopping processes to create the multiple Slater determinants,
that of exchanging a hole and a non-hole site is most probable, due to Large $U$,
and the number of these pairs of sites is maximized at $x=\frac{1}{2}$.
This causes insufficient description of the ground state by single determinant in LSNO at $x=\frac{1}{2}$.

Finally, we stress that the above stripe spin correlation
induced by the anisotropy of the transfer integrals is only seen under the checker-board type
charge order.
When the anisotropy $\delta=0.02$~eV is introduced to State S 
(see Sec.~\ref{sec_x_half_charge_order}) in the range  $0~{\rm eV}\le V \le 0.4$~eV,
the resultant state shows the spin order of N\'{e}el order type,
which is also seen in $\delta=0$ case in the range $0~{\rm eV}\le V \le 0.4$~eV.
The N\'{e}el order of half-filled $3z^2-1$-orbitals
appears under the homogeneous charge distribution and the small $\delta$ does not
affect the spin order.
Under the charge ordered condition, the single electron wavefunction is rather localized,
if not site-localized, then the transfer integrals are reduced effectively.
Then the small $\delta$ changes the spin order drastically into the spin stripe.

\section{Conclusion}\label{sec_conclusion}

We discussed the charge and spin order of LSNO ($x=\frac{1}{3}$) by using LSDA+U method
and those of LSNO ($x=\frac{1}{2}$) by using exact diagonalization of the double orbital
extended Hubbard model derived from LDA calculations.
In the exact diagonalization, charge and spin order is discussed 
by using charge-charge or spin-spin correlation functions. 
Excitation spectra of LSNO ($x=\frac{1}{2}$) are also calculated to show that 
the system is insulator.

In conclusion, LSNO with $x=\frac{1}{2}$ and $\frac{1}{3}$ are both insulator with charge and spin stripe order.
In both systems, diagonal hole stripes are separately located on Ni$^{3+}$ site in order to
reduce hole-hole interaction energy induced by inter-site Coulomb interaction $V$. 
We discussed the important role of multi-orbitals and mixing of multiple Slater 
determinants especially in high-doped $x=\frac{1}{2}$ system, where the charge order
is induced by the correlation energy of the inter-site Coulomb interaction $V$.
Charge order and the inter-site Coulomb interaction $V$ are directly related to 
the energy gap in the excitation spectra of the system with $x=\frac{1}{2}$.
Spin stripe occurs only under the condition of the existence of the charge order, 
with a help of anisotropy $\delta$ in diagonal hopping. 
Thus the spin stripe is determined by the electronic structure with smaller energy scale 
than that of the charge stripe. 
This is consistent with the observation of $T_{\rm CO}>T_{\rm SO}$.
Though the spin stripe is related to anisotropy, the mechanism is not of Jahn-Teller type 
unlike the usual ordering in transition metal oxides, 
because anisotropy in LSNO does not couple to the Jahn-Teller mode.

The stability of LNSO in different hole concentration $x$  
depends  sensitively upon several physical quantities, e.g. $t_{i\alpha j\beta}$, 
$\Delta$, $U$, $J$, $V$ and $\delta$, and 
we believe that we have successfully shown the new scope of 
the combination with the first principles electronic structure calculations 
and the many-electron theory. 
We have developed a very useful novel tool for the extremely large matrix 
of extended Hubbard Hamiltonian, the shifted-COCG method, which will be explain 
more details elsewhere.

Lastly we comment on layered cuprates. 
The band gap of  La$_2$CuO$_4$ (LCO) is 2~eV, 
narrower than La$_2$NiO$_4$, and hole doping makes systems metallic.~\cite{Wakimoto_1999}
Due to these facts, the screened Coulomb interaction in LCO becomes smaller than in LSNO
and the energy gain by hole hopping is more important in doped cuprates.  
This may be one of reasons why the hole stripe in
La$_{1.48}$Nd$_{0.4}$Sr$_{0.12}$CuO$_4$ runs in a direction along the n.n. pair 
and the hole concentration is one per two Cu sites in the stripe.~\cite{Tranquada_1995}

\begin{acknowledgments}

The authors are benefited greatly from discussions with M. Imada, Y. Tokura, and N. Nagaosa.
Calculations were done at the Supercomputer Center, Institute for Solid State 
Physics, The University of Tokyo and at Research Center for Computational
Science, Okazaki, Japan. 
This work was partially supported by a Grant-in-Aid for Scientific Research in Priority
Areas ``Development of New Quantum Simulators and Quantum Design'' (No.170640004) of
The Ministry of Education, Culture, Sports, Science, and Technology, Japan.

\end{acknowledgments}

\appendix
\section{Method to get smooth eigenvectors with respect to a parameter in Hamiltonian}
\label{app_cont}

We use Lanczos method and the inverse-iteration method with CG method alternatively
in order to obtain the exact eigenvectors.
These methods work well for getting ground states, but careful treatment
must be taken for solving two smooth eigenvectors with respect to $V$ 
in Sec.~\ref{sec_x_half_charge_order}.
Because both methods are the energy minimization process,
iterative application of them, with no care, gives the eigenvector of lower eigen energy
than that of the aiming eigen energy smoothly connected from the solved eigen energy
at adjacent values of $V$,
though one chooses the solved eigenvector at adjacent value of $V$ as a starting vector.
When the two levels are nearly degenerate, this problem becomes seriously important.
In the present case,
at least three energy levels cross with changing $V$ from 0~eV to 0.5~eV,
because at $V=0.5$eV, the lowest three levels are 
$36.7549$, $36.7551$, $36.7773$ in the unit of eV,
(the first one is a level of State S
and all the three levels are doubly degenerate).
State S locates at the $36.7939$~eV at $V=0.5$~eV, higher than these three levels,
while State S is the ground state at $V=0$~eV.

The first technique to avoid this problem is to use small number of
the dimension of the submatrix tridiagonalized by Lanczos method,
since Lanczos method has stronger tendency to reduce trial eigenvalue than that of inverse-iteration with CG method.
We tried successfully the dimension in the range 0 $\sim$ 80 depending on the case.
The second technique is to control the trial eigenvalue so that the residual is minimized and
the trial eigenvalue does not jump, in the whole processes.
Here, the residual is defined as $||(\hat{H}-(\mbox{trial energy}))(\mbox{trial vector})||$.
Because increase of the residual is a sign of the transition into lower level,
once it occurs, we dispose a new (and possibly lower) trial eigenvalue.

Lastly, the inner product among resultant eigenvectors of respective values of the specific
parameter ($V$ in the present case) must be checked.
The large value of inner product assures the smoothness of connection to
the resultant eigenvectors.
In the present case, 
the inner product between State S at $V=0$~eV and that at $V=0.5$~eV equals to 0.54 and
the inner product between State D at $V=0$~eV and that at $V=0.5$~eV equals to 0.73.
These values are large enough to assure the smoothness of these eigenvectors,
because no other eigenvectors at the end point $V=0.5$~eV can have larger inner product
than above respective states.

\section{Weight of the Slater determinants with complete charge ordered $V=0$}
\label{app_W_SDCCO}
One might have a question why State S has a finite weight 0.045 of SDCCO's at $V=0$, 
though the state has no charge order.
We conclude that the SDCCO's are not overweighted by following discussion.
Because the energy scale of the charge order is larger than that of
spin order in the present system, 
we neglect the spin configuration in each Slater determinant,
and count only the charge degrees of freedom.
Strong $U$ inhibits the configurations with site occupation more than two and zero 
and there exist $_8{\rm C}_4=70$ types of the charge configurations, 
where the site occupations are restricted to be one or two. 
A number of  charge configurations with complete checker-board type order equals to two.
Assuming that all configurations have the same weight due to the absence of charge order,
resultant weight of SDCCO's equals to $\frac{2}{70}\sim 0.029$,
which is consistent with above 0.045.

\section{Gap width in the bulk limit}\label{app_gap}

First we discuss the value of the energy gap  $0.9$~eV  in two way,
applying two extreme approximations, 
and show how inter-site Coulomb interaction opens the energy gap.
One of two approximations is such that  HIL and LAL are approximated 
by single particle excitation related to an extended hole/electron, 
and the other is such that they are approximated by a single particle excitation of
a site localized hole/electron.
In the actual picture, excitations are described by dressed quasi-particle/hole
and not by a single particle/hole, as is discussed later in the present section.

We begin with the case that HIL/LAL is approximated 
by single particle excitation of an extended state. 
Due to large splitting $\Delta=0.97$~eV between $3z^2-1$- and $x^2-y^2$-orbital, 
hole is doped into $x^2-y^2$-orbital. 
Therefore, low energy excitation mostly consists of $x^2-y^2$-orbital and 
we neglect $3z^2-1$-orbital in the present paragraph. 
The single electron wave functions of HIL and LAL belong to 
the same symmetry group labeled 
${\bm k}=(\pm\frac{\pi}{2},\pm\frac{\pi}{2})\frac{1}{a}$
as in Table \ref{tab_occ}. 
By unitary transformation, 
four plain waves labeled ${\bm k}=(\pm\frac{\pi}{2},\pm\frac{\pi}{2})\frac{1}{a}$
change the forms into four checker-board type wavefunctions
 $\cos(\frac{\pi}{2}x)\cos(\frac{\pi}{2}y)$,
 $\sin(\frac{\pi}{2}x)\sin(\frac{\pi}{2}y)$,
 $\cos(\frac{\pi}{2}x)\sin(\frac{\pi}{2}y)$,
 $\sin(\frac{\pi}{2}x)\cos(\frac{\pi}{2}y)$.
These four wavefunctions give the two types of charge distribution.
One type of them is shown in Fig.\ref{fig_charge_order}~(b) 
and the other is obtained with exchanging hole and non-hole sites. 
Two wavefunctions $\cos(\frac{\pi}{2}x)\cos(\frac{\pi}{2}y)$ and 
$\sin(\frac{\pi}{2}x)\sin(\frac{\pi}{2}y)$ do not interact 
with on-site Coulomb interaction $U$, 
because $(\cos^2(\frac{\pi}{2}x)\cos^2(\frac{\pi}{2}y)) (\sin^2(\frac{\pi}{2}x)\sin^2(\frac{\pi}{2}y))
=16\sin^2(\pi x)\sin^2(\pi y)=0$ on each lattice point. 
Instead, the interaction between them with inter-site Coulomb interaction $V$ 
is roughly estimated as $\frac{1}{n} \times \frac{1}{n} \times 4V \times n = \frac{4V}{n}$,
where $n=(\mbox{the number of sites whose amplitude is finite})=4$ in the present case.
This is occupied- and unoccupied-splitting. 
Assuming $\cos(\frac{\pi}{2}x)\cos(\frac{\pi}{2}y)$ is occupied,
then $\sin(\frac{\pi}{2}x)\sin(\frac{\pi}{2}y)$ is unoccupied and vice versa.
The same situation occurs for another pair of wavefunctions
$\cos(\frac{\pi}{2}x)\sin(\frac{\pi}{2}y)$ and $\sin(\frac{\pi}{2}x)\cos(\frac{\pi}{2}y)$.
Thus, the system becomes insulator with gap roughly estimated as $\frac{4V}{n}=0.5$~eV,
except $V=0$~eV where all the occupied- and unoccupied- states are degenerate 
(See Fig.\ref{fig_spectral_function_vs_V}(b)). 
Though the correlation induced by inter-site Coulomb interaction causes 
the splitting between occupied- and unoccupied- states (the energy gap),
this mechanism is different from the normal Mott insulator,
induced by on-site Coulomb interaction $U$.~\cite{State_S}
In addition to that, there is another difference between the current mechanism and
that of Mott insulator;
the wavefunction of HIL and LAL have checker-board type charge order,
not sharing the center of charge distribution in the current mechanism,
unlike they shares the center of charge distribution in Mott insulator.
However, there is a problem with this estimation of the gap.
Following the present estimation, energy gap vanishes in the bulk limit,
because the gap  is inversely proportional to the system size.

In the next estimation, HIL/LAL excitation is approximated by a single particle
excitation of site localized hole/electron.
Neglecting hopping integrals, the ground state of the present Hamiltonian
consists of four Ni$^{2+}$ (spin triplet, u$^1$v$^1$) and four doublet Ni$^{3+}$
(spin doublet, u$^1$) located on respective sublattice,
where u and v denote $3z^2-1$- and $x^2-y^2$- orbital respectively.
All these components are assigned to the Slater determinants
with complete checker-board type charge order (SDCCO's) in Sec.~\ref{sec_x_half_charge_order}.
Then the LAL is $x^2-y^2$-orbital of Ni$^{3+}$
and  the HIL is $x^2-y^2$-orbital of Ni$^{2+}$ (both have a majority spin).
Consequently, the energy gap is estimated
as $\{[E({\rm Ni}^{2+})]+8V-[E({\rm Ni}^{3+})+4V]\}
-\{E({\rm Ni}^{2+})-E({\rm Ni}^{3+})\}=4V=0.5$eV, 
where $E(\cdot)$ denote the ground state energy of Ni ion for each
ionization state.
This is an exact solution even in the bulk limit.

The result of the present calculation is 
in the intermediate region of above two estimations;
the calculated energy gap $0.9$~eV
is greater than $0.5$~eV in extreme cases by the extended HIL/LAL approximation
and less than $2.0$~eV by the site localized HIL/LAL approximation.
Therefore, in the calculated excitations 
(with respect to State D at $V=0.5$~eV in Sec.~\ref{sec_x_half_charge_order})
are well described by 
the wavefunction in intermediate region between site localized and extended states.
And  the split between occupied- and unoccupied- states occurs
due to long-ranged (inter-site) Coulomb interaction,
though they are {\it not sharing the center} and does not have large overlap.

Finally, we discuss on the bulk limit of the energy gap in the present calculation.
Taking the limit such that transfer integrals go to infinity,
the system becomes paramagnetic metal.
Taking the limit such that transfer integrals go to zero
the system becomes antiferromagnetic insulator.
Therefore, there exists a critical transfer integral
where the bulk limit of the system changes from insulator (metal) to metal (insulator).
The high weight of the Slater determinant with complete checker-board type charge order
($>0.4$ at $V=0.5$~eV, as in Fig.\ref{fig_weight_SDCCO}) strongly suggests that
the present choice of transfer integrals makes the bulk limit insulator,
and that is consistent with the real LSNO ($x=\frac{1}{2}$).



\begin{thebibliography}{999}
\bibitem{LDA-result} The calculated ground state by the LDA result for $x=0$ is the paramagnetic metal without orbital polarization.
\bibitem{State_S} The ground state in the range $V<0.41$eV (State S) is 
insulator with no charge order,
such that the charge gap opens due to on-site Coulomb interaction $U$.
\end{thebibliography}
\end{document}